\documentclass[preprint2]{aastex}

\newcommand{\simgt}{\,\rlap{\lower 3.5 pt \hbox{$\mathchar \sim$}} \raise
1pt \hbox {$>$}\,}
\newcommand{\simlt}{\,\rlap{\lower 3.5 pt \hbox{$\mathchar \sim$}} \raise
1pt \hbox {$<$}\,}

\shorttitle{The DM distribution function}
\shortauthors{Hansen \& Sparre}

\begin{document}


\title{A derivation of (half) the dark matter distribution function}


\author{Steen H. Hansen \& Martin Sparre}
\affil{Dark Cosmology Centre, Niels Bohr Institute, University of Copenhagen,\\
Juliane Maries Vej 30, 2100 Copenhagen, Denmark}
\email{hansen@dark-cosmology.dk, sparre@dark-cosmology.dk}



\begin{abstract}
  All dark matter structures appear to follow a set of universalities,
  such as phase-space density or velocity anisotropy profiles,
  however, the origin of these universalities remains a mystery.  Any
  equilibrated dark matter structure can be fully described by two
  functions, namely the radial and the tangential velocity
  distribution functions (VDF), and when we will understand these two
  then we will understand all the observed universalities. Here we
  demonstrate that if we know the radial VDF, then we can derive and
  understand the tangential VDF. This is based on simple dynamical
  arguments about properties of collisionless systems. We use a range
  of controlled numerical simulations to demonstrate the accuracy of
  this result.  We therefore boil the question of the dark matter
  structural properties down to understanding the radial VDF.
\end{abstract}




\section{Introduction}

A growing number of seeming universalities have been identified in
numerical simulations of dark matter structures. Most of these are
integrated quantities, such as the density profile \citep{nfw,moore}, the
pseudo phase-space density \citep{taylornavarro}, and the velocity
anisotropy \citep{hansenmoore}.  The cause of these universalities
remains, however, essentially unknown.

The origin of the universalities may lie in some fundamental property
of dark matter, be it some statistical mechanics
\citep{lyndenbell,hjorthwilliams} or optimization of some generalized
entropy \citep{plastino,zurichstudents,2010MNRAS.406.2678H,2012MNRAS.419.1667H}. It may also be associated to
dynamical effects, like radial orbit instability
\citep{henriksen2009,bellovary}, or phase mixing or violent relaxation
\citep{lyndenbell, kandrup}.  Alternatively, it could just be a
``coincidence'', since all structures have been built up through similar
processes of mergers and accretion \citep{gsmh,smgh07}.

A first step towards answering the question of the origin of the
integrated universalities, is to look at the actual distribution of
velocities. Also the actual shape of the velocity distribution
function (VDF) has been suggested to be universal \citep{hansenzemp},
which naturally could explain all the integrated universalities.

Asking the question about what dark matter structures fundamentally
want, is different from asking what dark matter structures in an
expanding universe actually end up doing. We will therefore not be
considering structures from cosmological simulations, since their
profiles often have merger history and environment dependent
profiles. We will instead consider a range of numerical simulations
where we have better control of their evolution. In this way we can
repeatedly perturb the structures in controlled manners, as well as
giving the structures sufficiently time that phase mixing between
individual perturbations may be more complete than what is the case in
cosmological simulations.

A non-trivial dark matter VDF also has direct implications for direct
dark matter experiments (see e.g. 
\cite{vergados,2009JCAP...01..037F,2010JCAP...02..030K}
for discussions and references).

We here present numerical evidence that the origin of the shape of the
tangential VDF is simple dynamics, hence supporting the idea that dark
matter wants to follow very simple dynamical rules. This explains the
origin of the velocity anisotropy profile in the inner region of dark
matter structures, with no seeming need for advanced statistical
mechanical or generalized entropic principles.  However, as we will
point out, we are still left with an unknown origin of the radial VDF
and hence also the density profile and the pseudo-phase space density
profile are not explained yet.

Below we will explain the surprisingly simple dynamical reason for the
full shape of tangential part of the VDF, and we will perform
numerical simulations supporting this conclusion.

{Some of these physical arguments have been presented previously
\citep{2009ApJ...694.1250H}, however, the simulations presented here are
significantly improved. In particular we create a set of controlled
perturbations using energy exchange reminiscent of violent relaxation
and dynamical friction, which allows the particle distributions to
change significantly, without having the structure depart from
spherical symmetry. At the same time the structures are analysed only
after convergence to a fully stable configuration has been achieved.}

\section{Decomposition}
Let us consider a particle moving in the smooth and spherical
potential of many collisionless particles.
The velocity of this particular particle can be decomposed into three
components, namely the radial and the two tangential components.
With such decomposition we can consider all particles in a given radial
bin, and get the velocity distribution function (VDF) in both the
radial and tangential directions. If the structure is non-rotating, then
the two tangential VDF's will be identical. 

It has long been known that the radial and the tangential VDF's are
different, and physically this difference may seem very reasonable for
equilibrated systems, as we will now explain. 

{We will first discuss the radial VDF.}  Consider a thin spherical
bin, at radius $r$.  If we consider the velocity components moving
outwards in the radial direction, then those must be compensated by
particles further out moving inwards. This compensation must depend on
the particular density profile of the structure.  This is most clearly
seen in the Eddington inversion method \citep{eddington,binneytremaine}, from
which one can easily derive the full radial VDF from the full density
profile. 

%The Eddington method only works for
%the case of $\beta=1 - \sigma^2_{\rm tan}/\sigma^2_{\rm rad} = 0$,
%which implies that the radial and tangential VDF's are identical.

Now, let us instead consider the tangential velocity
components. Instantaneously the components are moving in the
tangential plane. {For particles with the circular speed this
  means that the component is moving in constant density and constant
  potential.  Particles moving slowly in the tangential plane will
  still be near the same density and potential after a short time
  interval.}  That implies that the equilibrium can be achieved simply
by having other components in the same radial bin moving in the
opposite direction. Therefore, whereas the radial VDF's depend on the
full radial density profile, then the tangential VDF's apparently do
not have to concern themselves with other radial bins.  The tangential
VDF could therefore, in principle, be the same at all radii.

This argument only holds instantaneously. {Particles whose
  tangential velocity components are high, will after a longer time
  interval, $\Delta t$, be moving outwards into lower density regions,
  and will therefore later have converted their tangential component
  into both radial and tangential velocity components.} Effectively
this implies that the argument, that the tangential velocity component
moves in constant potential and constant density, only holds for the
low velocity particles. Below we will use numerical simulations to
show that effectively the breakdown of this argument happens around
$70\%$ of the escape velocity, $v_{\rm esc}$.

\subsection{Low velocity component}
\label{sec:explain}
We have argued above that the low velocity component of the tangential
VDF should have a simple shape, which should be the shape of
collisions-less particles moving in a constant potential and constant density.
To simulate a uniform medium is rather difficult using N-body simulations, 
because any power or noise
will induce gravitational collapse, which leads to a departure from 
homogeneity. Instead one can make a very simple analytical argument, which
allows one to derive the VDF of a homogeneous medium.

Let us consider a spherical structure which has a known density
profile, $\rho(r)$.  If one has $\beta =0$, then we can use Eddington's
method to derive the VDF, 
\begin{equation}
f(E) = \frac{1}{\sqrt{8} \pi^2} \int_0 ^E 
\frac{d^2\rho}{d \Psi^2} \frac{d \Psi}{\sqrt{E-\Psi}} \, ,
\end{equation}
where $\Psi(r)$ is the relative potential as a function of radius, and
$E$ is the relative energy, $E =\Psi -mv^2/2$. 
Eddington's method provides the unique ergodic distribution
function \citep{binneytremaine}. 
We use this method to find the VDF at all radii.

Imagine that the structure is
particularly simple, namely with a constant density slope, $\gamma =
d{\rm log} \rho/d {\rm log} r$, over a very large radial range.  To be
concrete, say that $\gamma=-2$ over 20 orders of magnitude in radius,
and then truncated abruptly inside and outside this range. Eddington's
method shows us that this (isothermal sphere) has a VDF which is a
Gaussian at all radii (except for the details arising from the
truncation).

{Now, consider a more shallow slope, e.g. $\gamma = -1.2$ or
$-0.4$. For any value of $\gamma$ we can use Eddington's method to
calculate the VDF, and at any radius it will have exactly the same
shape as function of $v_r/\sigma_r$. 
For each density slope we use this method to 
find the unique distribution function.

Finally we can extrapolate this
approach to $\gamma=0$, which is identical to the case of constant
density and constant potential. The shape of the VDF for this case is
\begin{equation}
f(v) \sim \left(1 +  \frac{v^2}{3 \sigma^2}\right) ^{-5/2} \, .
\label{eq:ftan}
\end{equation}
$\gamma=0$ is the condition for a particle moving in constant density
and potential, and this is therefore the shape for the tangential VDF.
In the case with $\gamma=0$ there is no difference between the radial
and tangential directions, and the assumption of $\beta=0$ in the
derivation is therefore correct, and the technique is thus self
consistent.}

We now have an analytical expression for the shape of the tangential
VDF at small velocities, and only the normalizations are
unknown. These are the overall normalization (which must be related to
the density at that radius), and the normalization of the velocity
(which is related to the tangential velocity dispersion).

\subsection{High velocity component}

The high velocity components are possibly even simpler to
describe. {If a velocity component is purely tangential at a given
  time, then shortly later it will be a combination of tangential and
  radial (unless it happens to have exactly the circular speed)}. We
should therefore expect that the shape of the tangential and radial
components are similar at high velocities.

There is only one complication, namely the normalization. The overall
normalization must be identical between the radial and tangential
components (since this is just the local density), however, as opposed
to the low velocity component discussed above, the normalization of
the high velocities components must be absolute, i.e. $v_{\rm rad} \sim
v_{\rm tan}$.

%and not normalized by their respective dispersions,
%$v_{\rm rad}/\sigma_{\rm rad} \sim v_{\rm tan}/\sigma_{\rm tan}$.

\subsection{The transition}

The transition from the {\em low} to the {\em high} velocity
components in a real system must be smooth, and is probably rather
non-trivial, however, for simplicity we will here make the
approximation that the transition is abrupt, and we make no attempt to
make it smooth.  Practically, we simply assume that the transition
always happens near $0.7\, v_{\rm esc}$. We will address this issue
further in the discussion section.

\section{Simulations}

The first simulation is a cold collapse, where the inclusion of substructure
breaks the spherical symmetry.
We distributed $5 \times 10^5$ particles according to a Hernquist
density profile~\citep{hernquist} with scale radius $1$, and a cutoff
at $200$. In addition $5 \times 10^5$ particles, with the same mass as
the main halo particles, were distributed in $24$ identical subhaloes,
also having Hernquist density profiles, but with a scale radius of
$0.5$ and a cutoff radius of $5$. The centers of the subhaloes were
sampled in the same way as the particles in the main halo. The
velocities of all the particles were initially zero. The total mass in
the simulation was $1$. We ran the simulation for 200 time units,
which corresponds to 200 dynamical times at the scale radius for the
initial structure.  Such a cold collapse is similar to the simulations
by van Albada (1982).

For all the non-cosmological simulations discussed here, we used the
parallel N-body simulation code, Gadget2 \citep{springel}. For further
details on the cold collapse see \cite{sparrehansen}.

\begin{figure}[thb]
	\centering
	\includegraphics[angle=0,width=0.48\textwidth]{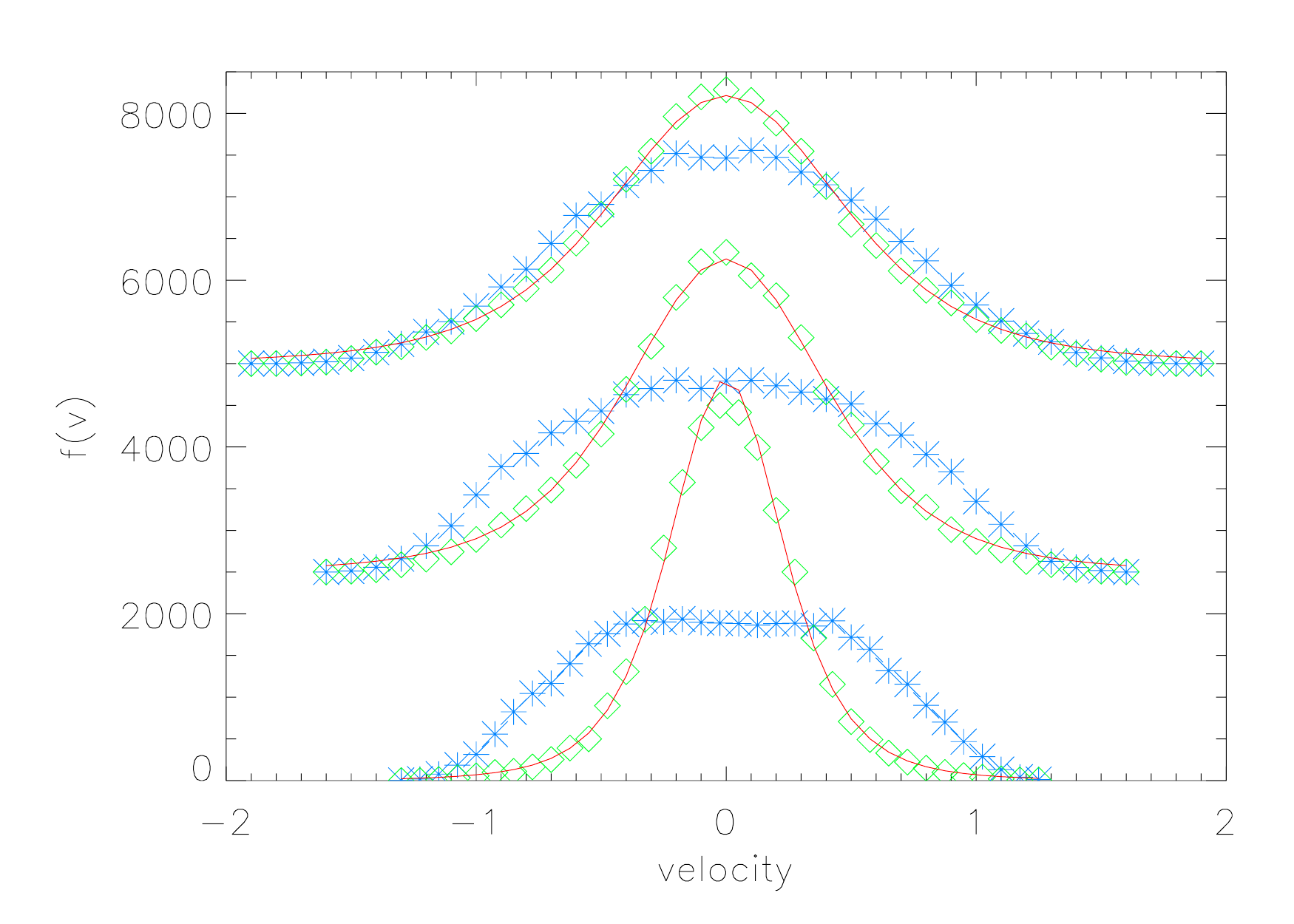}
	\caption{{Radial (blue stars) and tangential (green diamonds)}
          VDF for three radial bins in the cold collapse
          simulation. From top to bottom the radial bins are at
          $\gamma = -1.6, -2.0$ and $-2.4$, and the bins are shifted
          vertically to improve readability. The red (solid) lines are
          {all of the same theoretical shape in
            eq.~(\ref{eq:ftan})}, which are seen to provide an
          acceptable fit in the low velocity region.}
\label{fig:infall.lin.lin}
\end{figure}

To sample the VDF's we distribute the particles in radial bins with
the same number of particles in each bin. In
Figure~\ref{fig:infall.lin.lin} we show both the radial (blue stars)
and tangential (green diamonds) VDF for three radial bins, chosen near a
slope of $\gamma = -1.6, -2.0$ and $-2.4$ (from top to bottom).  We
also show the predicted shape of the tangential VDF (solid line),
which is clearly seen to provide an acceptable fit in the low velocity
region. It is also clear that the radial and the tangential VDF's are
very different in the low velocity region.

\begin{center}
\begin{figure}[thb]
	\includegraphics[angle=0,width=0.48\textwidth]{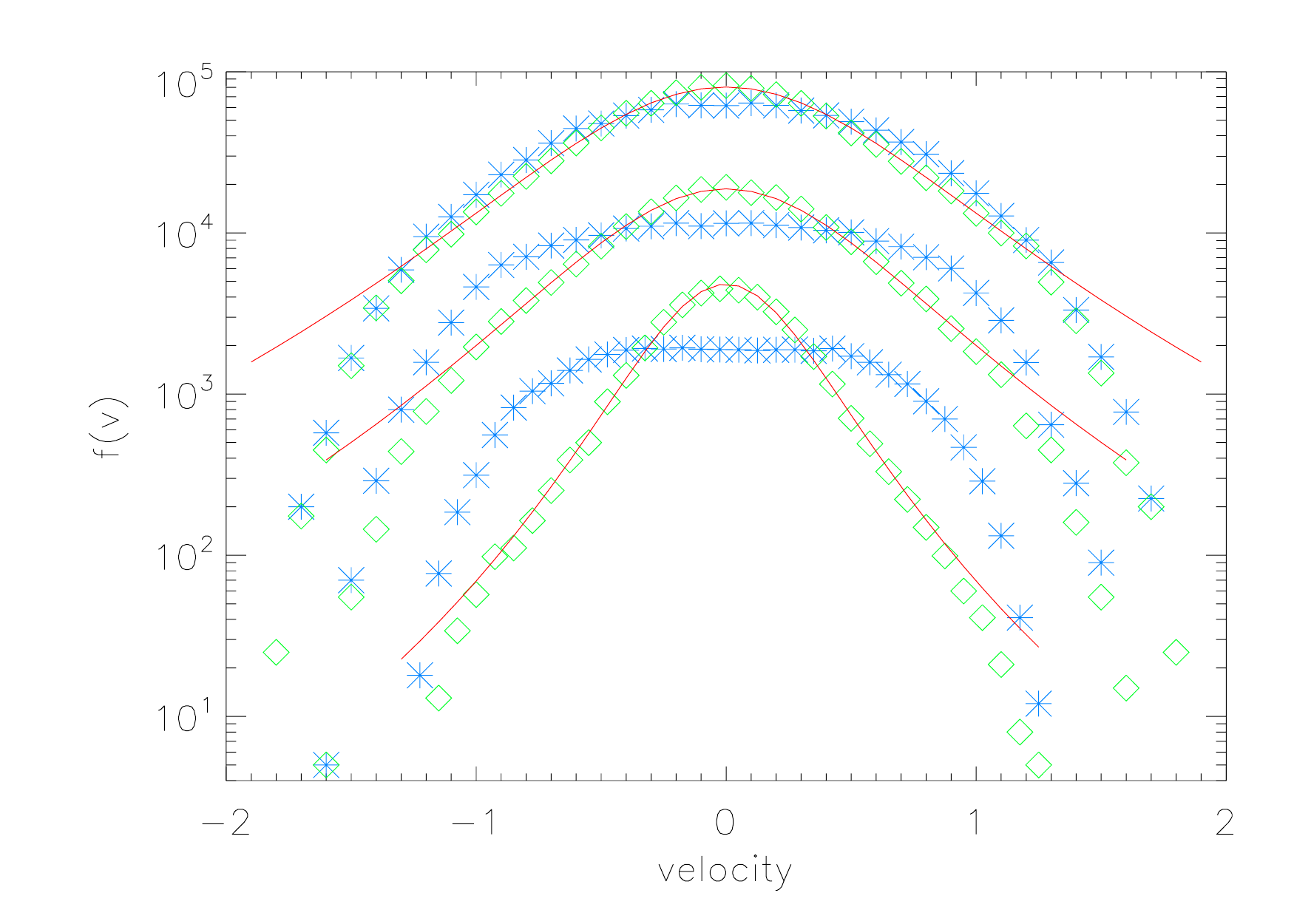}
	\caption{{Same as figure 1, only linear-log to see details of
          the high-energy tail.  Radial (blue stars) and tangential
          (green diamonds) VDF }for three radial bins in the cold
          collapse simulation. From top to bottom the radial bins are
          at $\gamma = -1.6, -2.0$ and $-2.4$, and the bins are
          shifted vertically to improve readability. The red (solid)
          lines are of the theoretical shape for the low velocity
          region.  For the high velocities it is clearly seen that the
          radial and the tangential VDF's approach each other
          rapidly.}
\label{fig:infall.lin.log}
\end{figure}
\end{center}

To see the details of how the radial and tangential VDF's start
agreeing in the high velocity region we plot the VDF's from the same three
radial bins in a lin-log space in Figure~\ref{fig:infall.lin.log}.  It
is clear that for high velocity particles the radial VDF's (blue
stars) are in good agreement with the tangential VDF's (green
diamonds). Interestingly, the tangential VDF's can be approximated
with the theoretical solid curve for low energies, and with the radial
VDF's for high velocities. And for the radially innermost bins (at
slopes shallower than $-2$) this transition happens to be just around
$v = 0.7 \, v_{\rm esc}$.  Only for radial bins further out than
$\gamma = -2$, it seems needed to use a more smooth transition.

\subsection{G-perturbations}

In order to test further the theoretical claims for the tangential
VDF, we wish to construct a perturbation/equilibration scheme, which
allows the VDF's to change significantly, without having the structure
depart from spherical symmetry.

We set up structures in perfect equilibrium. These structures may have
any density profile, and have zero anisotropy or follow and
Osipkov-Merritt beta profile \citep{binneytremaine}. Now we increase
the value of the gravitational constant by $20\%$. This increases the
potential and makes the structure contract, and after a few dynamical
times a new equilibrium is reached. Next we repeatedly increase or
decrease the gravitational constant, and between each change we allow
the structure to phase-mix and find a new equilibrium.  After 20 such
perturbations we use the standard value of $G$, and let the structure
relax completely.  For further details, see ~\cite{sparrehansen}.

\begin{figure}[thb]
	\centering
	\includegraphics[angle=0,width=0.48\textwidth]{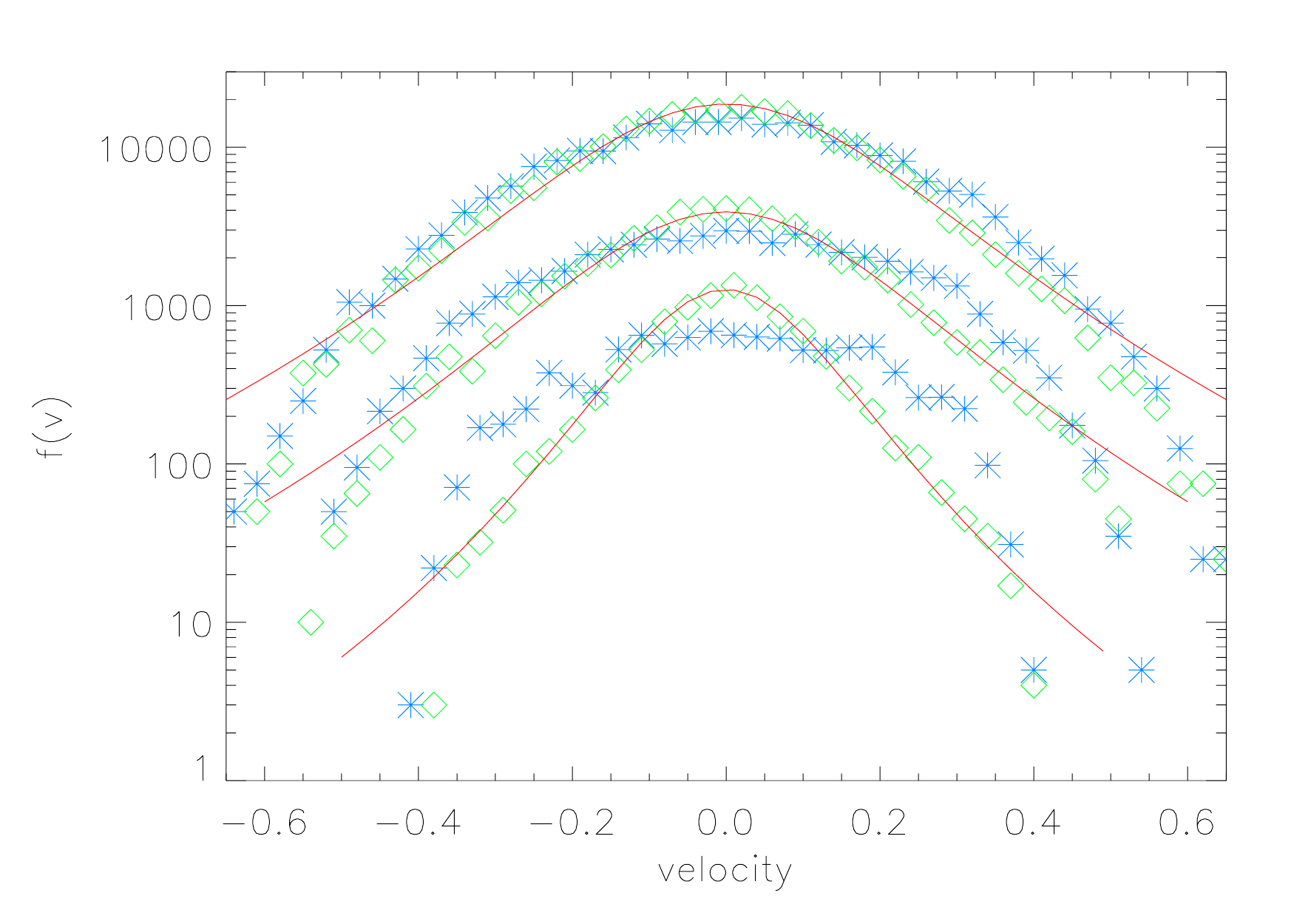}
	\caption{Radial and tangential VDF for three radial bins in
          the G-perturbation simulation. From top to bottom the radial
          bins are at $\gamma = -1.8, -2.3$ and $-3$, and the bins are
          shifted vertically to improve readability. {All the red
            (solid) lines are of the same theoretical shape in
            eq.~(\ref{eq:ftan})} for the low velocity region.  For the high
          velocities it is clearly seen that the radial and the
          tangential VDF's are very similar.}
\label{fig:omg19.lin.log}
\end{figure}

In Figure~\ref{fig:omg19.lin.log} we show three radial bins from a
simulation which initially was constructed as an isotropic Hernquist
profile. This in particular means that the initial conditions (before
the G-perturbations were executed) had identical radial and
tangential VDF's. Now, after the perturbations and subsequent
relaxation there is a large difference between the radial and
tangential VDF for small velocities. Instead, at high velocities the
tangential and radial VDF quickly approach one another. As is also
visible from the figure, the tangential VDF is well fitted by the
theoretical prediction for small velocities.

\subsection{Explicit energy exchange}

Collisionless particles experience different kinds of energy exchange
between each other, in particular through violent relaxation (where
the changing potential implies that the particle energies change), and
through dynamical friction (which transfers energy from the fast to
the slower particles).  We therefore consider a perturbation where the
spherical symmetry is again conserved, however, we allow the particles
to exchange energy amongst each other. This is done in such a way,
that each radial bin conserves energy, whereby both density and
dispersion profiles are unaffected by the perturbation itself. 
{This energy exchange is instantaneous and the subsequent evolution
is with normal collisionless dynamics.}
After
each perturbation we again allow for sufficient phase mixing
\cite[see][for details]{hjs2010}. After sufficiently many
perturbations (typically 20 or 30) the structures have converged to a
stable state, which will not change when exposed to further similar
perturbations \citep{hjs2010, barber2012}.

\begin{figure}[thb]
	\centering
	\includegraphics[angle=0,width=0.48\textwidth]{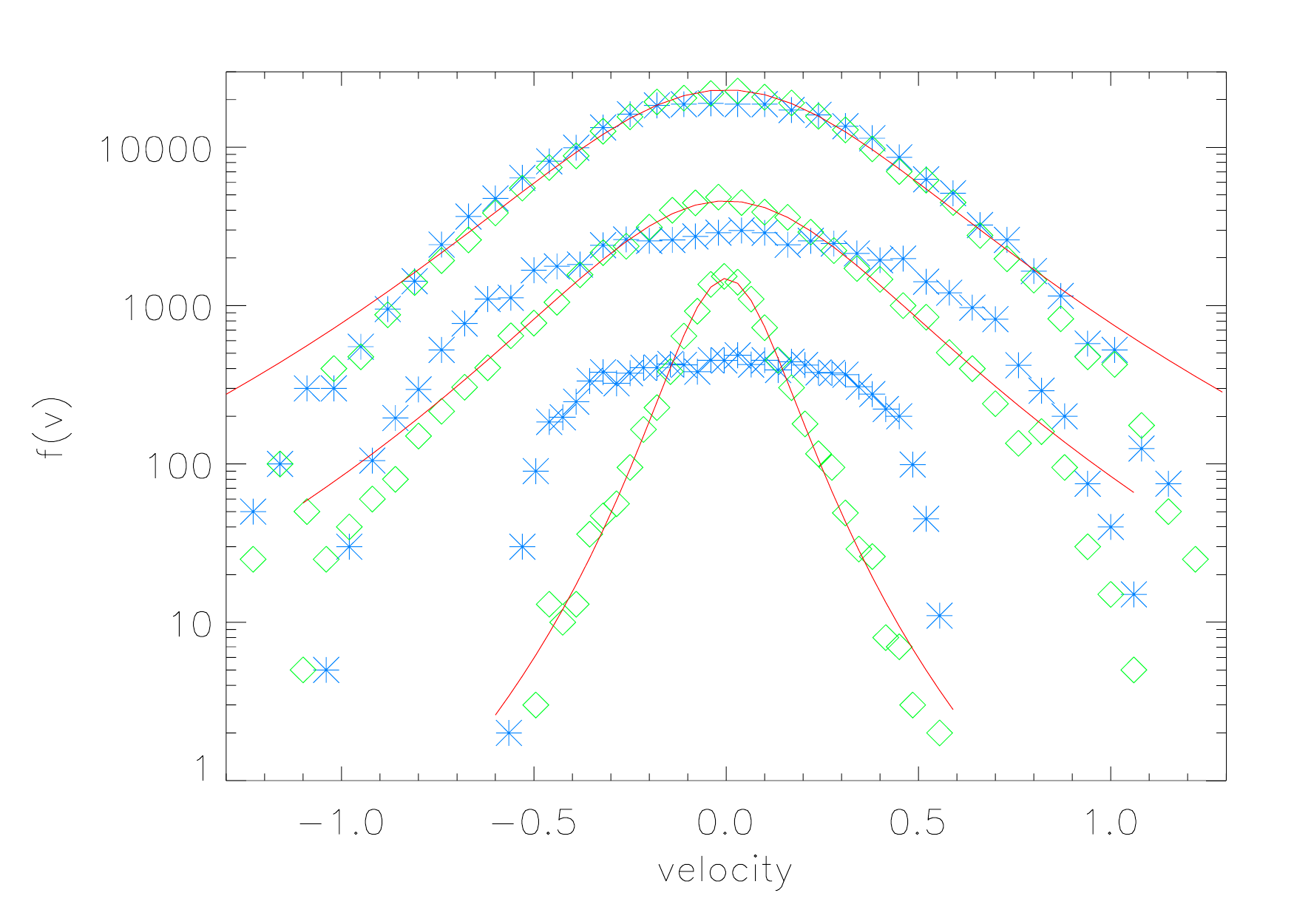}
	\caption{Radial and tangential VDF for three radial bins after
          perturbations by explicit energy exchange. From top to
          bottom the radial bins are at $\gamma = -1.7, -2.4$ and
          $-3.0$, and the bins are shifted vertically to improve
          readability. The red (solid) lines are of the theoretical
          shape for the low velocity region.  For the high velocities
          it is clearly seen that the radial and the tangential VDF's
          rapidly approach each other.}
\label{fig:hjs035.lin.log}
\end{figure}

In Figure~\ref{fig:hjs035.lin.log} we present the VDF from three radial
bins in the final structure, which are taken at density slopes of
$\gamma = -1.7, -2.4$ and $-3.0$. Again we see that the final VDF's
agree well between the radial and tangential for high velocity 
components, and that the low velocity components are well fitted
by the analytical expression.

\section{Discussions}

We have demonstrated that three extremely different artificial and controlled 
perturbations all lead to a tangential VDF, which is in good agreement
with the theoretical prediction. This result is in good agreement with
earlier studies including head-on collisions and galaxy formation
\citep{hansenzemp,2009ApJ...694.1250H}, and provides very strong evidence that
the origin of the shape of the dark matter tangential VDF is indeed
as simple as explained in section~\ref{sec:explain}. An important difference from 
earlier studies is
that we have here been investigating structures which have been exposed
to controlled perturbations, and analysed only after convergence to
a stable configuration has been achieved.

Let us remind the idea behind this paper.  If we know the radial and
the tangential VDF, then we know everything else, such as the
phase-space density, the velocity anisotropy and the density profiles.
For some idealized structures it is possible to derive the radial VDF
directly from the density profile, and our results here imply that in
that case we can derive the tangential VDF. That is, {\em if} we have
the radial VDF, {\em then} we can derive the tangential VDF.

\begin{figure}[thb]
	\centering
	\includegraphics[angle=0,width=0.48\textwidth]{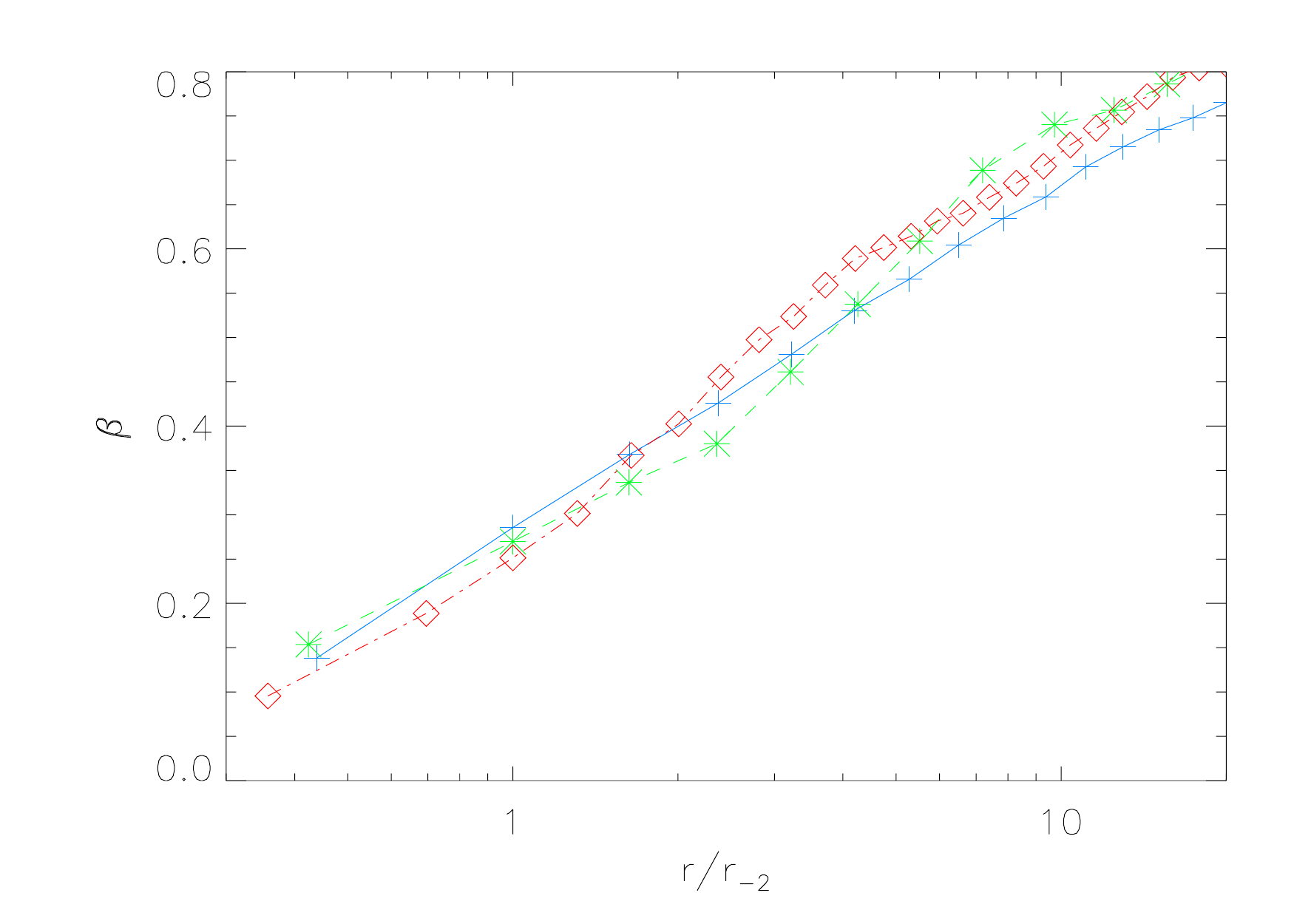}
	\caption{{The radial variation of $\beta$ as function of
            radius for the structures considered in figures 1-4
            (solid line HJS perturbation, dashed line G-perturbation,
            dot-dashed line infall simulation). The structures all
            have a strong variation, going from essentially isotropic
            towards the central region, to radial orbits in the outer
            regions.}}
\label{fig:beta}
\end{figure}

In the derivation of the tangential VDF we made no assumptions about
the anisotropy of the system, and the tangential VDF should therefore
be the same for all systems and at all radii, irrespective of their
anisotropy profiles. 
This statement naturally only holds for realistic systems which
  have been perturbed and allowed to relax. One can always create
  systems in a quasi equilibrium states which may even have highly
  different distribution function. Systems created away from an
  equilibrium state most often also have very different tangential
  distribution functions. However, as we are demonstrating in this
  paper, all systems which are exposed to sufficient perturbations and
  subsequently allowed to relax to a quasi equilibrium state, will
  indeed have a tangential VDF of exactly this shape.
In figure \ref{fig:beta} we present the
anisotropy profiles for the 3 systems considered. In particular one
sees that the anisotropy has a strong radial variation, going from
essentially isotropic in the central region, to radially dominated
orbits in the outer regions. And yet the shape of the tangential VDF
is the same at all radii.

It has previously been suggested that possibly the radial VDF is
sufficiently close to a rescaled VDF resulting from the Eddington
method \citep{2009ApJ...694.1250H}. We have tested this suggestion by fitting
the density profile and then using the Eddington method to extract the
radial VDF at all radii. However, the resulting VDF is not an 
accurate representation of the actual radial VDF. This means that
we are still not at the point of understanding the radial VDF.

\begin{figure}[thb]
	\centering
	\includegraphics[angle=0,width=0.48\textwidth]{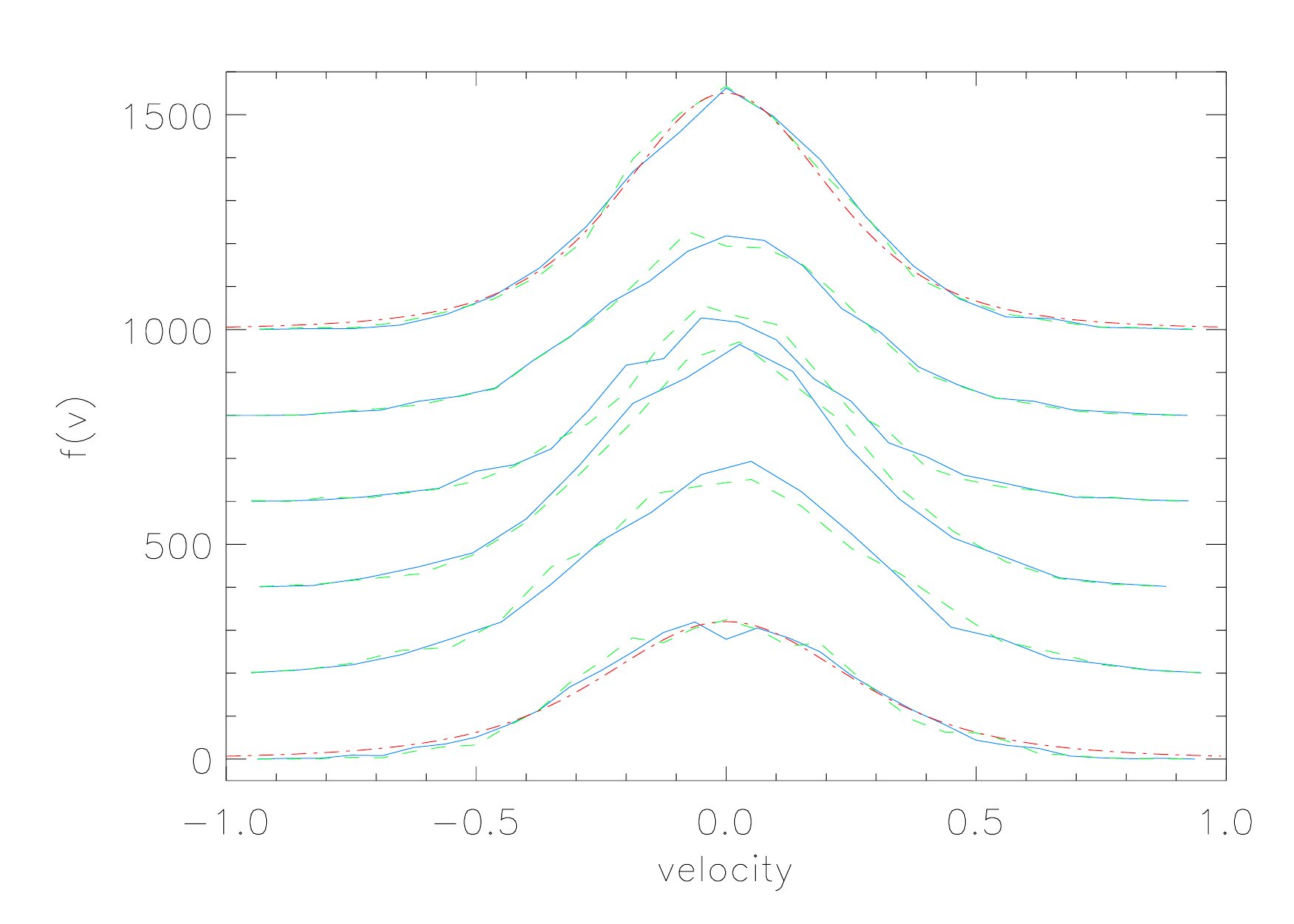}
	\caption{Radial (solid) and tangential (dashed) VDF for an
          inner bin (but not including approximately 35.000 of the
          most central particles). It is clear, that for all the
          structures the radial and tangential VDF are very
          similar. We add the same theoretical shape of
            eq.~(\ref{eq:ftan}) in red (dot-dashed) lines, just to
            demonstrate the agreement. The structures included here
          are the cold collapse, two different G-perturbations and
          three different HJS2010 perturbations, covering a range of
          initial density and anisotropy profiles, {in particular
            do we here present results for structures which were
            initially set up with a shallow central density
            profile}. Velocities are scaled by the escape velocity and
          shifted vertically to improve readability. For this figure
we selected structures which were created with zero inner slope,
$\gamma_{\rm initial}=0$, before perturbations were applied.}
\label{fig:central}
\end{figure}

One interesting aspect of the arguments presented for the shape of the
tangential VDF is, that for $\gamma = 0$ it should be identical to the
radial one.  That trend is already clear from the figures, namely that
the tangential VDF is suggestively close to the radial one for the
innermost bins (upper curves in all figures).  To test this further,
we selected a radial bin in the inner region (outside of 5 times the
softening length for each structure, and also outside a further 30.000
particles). The result is seen in Figure~\ref{fig:central}, where the tangential
(dashed) and radial (solid) VDF's are seen to be very similar. The
local density slope at these bins were ranging from $\gamma = -0.4$ to
$-1$ (bottom to top on Figure~\ref{fig:central}).

We have discussed that the transition between high and low velocity
may be approximated as a rather sharp transition. This is certainly a
good approximation for the inner region (inside a slope of $\gamma =
-2$). However, at larger radii it is clear, that a more smooth
transition would provide a more accurate representation of the
tangential VDF. For all the radial bins and for all the structures
considered in this paper, we have estimated the best velocity for the
transition, $ v_{\rm trans}$, and it appears that this transition is
in the range $0.6 \, v_{\rm esc} < v_{\rm trans} < 0.8 \, v_{\rm
  esc}$.  Thus, in order to avoid unnecessary fitting parameters, it
is a rather good approximation to fix this at $v_{\rm trans} = 0.7 \,
v_{\rm esc}$.

\section{Conclusions}
We have demonstrated that {\em half} of the distribution function
(specifically, the tangential velocity distribution function)
%, $f(v_{\rm tan}, r)$), 
for dark matter structures can be understood from simple
dynamical arguments. This implies that when we will eventually be able
to derive the {\em other half} (namely the radial velocity
distribution function), 
%$f(v_{\rm rad},r)$) 
then we understand all the properties of dark
matter structures, including the seeming universalities of the
density, phase-space density and velocity anisotropy profiles.

We saw that the derivation of the tangential VDF
%, $f(v_{\rm tan},r)$,
did not require any reference to statistical mechanics or generalized
entropy, but instead appears as a result of very simple dynamics.  It
now remains to derive the radial VDF, 
%$f(v_{\rm rad},r)$, 
and it will be
interesting to see if this will also be possible based on similar
basic dynamical arguments.

\medskip

\noindent
{\bf Acknowledgements}\\
The Dark Cosmology Centre is funded by the Danish National Research
Foundation.  The simulations were performed on the facilities provided
by the Danish Center for Scientific Computing.

%\noindent {\small Correspondence should be addressed to S.H.H\\ (e-mail: hansen@dark-cosmology.dk).}

\end{document}